\documentclass[twocolumn,showpacs,preprintnumbers,amsmath,amssymb,prl]{revtex4}
\usepackage{dcolumn}
\usepackage{bm}
\usepackage[dvips]{graphicx}
\usepackage{wrapfig}
\usepackage{psfrag}
\begin{document}

\def\sh{\mathop{\rm sh}\nolimits}
\def\ch{\mathop{\rm ch}\nolimits}
\def\var{\mathop{\rm var}}\def\exp{\mathop{\rm exp}\nolimits}
\def\Re{\mathop{\rm Re}\nolimits}
\def\Sp{\mathop{\rm Sp}\nolimits}
\def\kp{\mathop{\text{\ae}}\nolimits}
\def\bk{{\bf {k}}}
\def\bp{{\bf {p}}}
\def\bq{{\bf {q}}}
\def\lra{\mathop{\longrightarrow}}
\def\Const{\mathop{\rm Const}\nolimits}
\def\sh{\mathop{\rm sh}\nolimits}
\def\ch{\mathop{\rm ch}\nolimits}
\def\var{\mathop{\rm var}}
\def\mK{\mathop{{\mathfrak {K}}}\nolimits}
\def\mR{\mathop{{\mathfrak {R}}}\nolimits}
\def\mv{\mathop{{\mathfrak {v}}}\nolimits}
\def\mV{\mathop{{\mathfrak {V}}}\nolimits}
\def\mD{\mathop{{\mathfrak {D}}}\nolimits}
\def\mN{\mathop{{\mathfrak {N}}}\nolimits}
\def\mS{\mathop{{\mathfrak {S}}}\nolimits}

\def\Re{\mbox {Re}}
\newcommand{\Z}{\mathbb{Z}}
\newcommand{\R}{\mathbb{R}}
\def\mK{\mathop{{\mathfrak {K}}}\nolimits}
\def\mR{\mathop{{\mathfrak {R}}}\nolimits}
\def\mv{\mathop{{\mathfrak {v}}}\nolimits}
\def\mV{\mathop{{\mathfrak {V}}}\nolimits}
\def\mD{\mathop{{\mathfrak {D}}}\nolimits}
\def\mN{\mathop{{\mathfrak {N}}}\nolimits}
\newcommand{\ccm}{{\cal M}}
\newcommand{\cE}{{\cal E}}
\newcommand{\cV}{{\cal V}}
\newcommand{\cI}{{\cal I}}
\newcommand{\cR}{{\cal R}}
\newcommand{\cK}{{\cal K}}
\newcommand{\cH}{{\cal H}}

\def\br{\mathop{{\bf {r}}}\nolimits}
\def\bS{\mathop{{\bf {S}}}\nolimits}
\def\bA{\mathop{{\bf {A}}}\nolimits}
\def\bJ{\mathop{{\bf {J}}}\nolimits}
\def\bn{\mathop{{\bf {n}}}\nolimits}
\def\bg{\mathop{{\bf {g}}}\nolimits}
\def\bv{\mathop{{\bf {v}}}\nolimits}
\def\be{\mathop{{\bf {e}}}\nolimits}
\def\bp{\mathop{{\bf {p}}}\nolimits}
\def\bz{\mathop{{\bf {z}}}\nolimits}
\def\bbf{\mathop{{\bf {f}}}\nolimits}
\def\bb{\mathop{{\bf {b}}}\nolimits}
\def\ba{\mathop{{\bf {a}}}\nolimits}
\def\bx{\mathop{{\bf {x}}}\nolimits}
\def\by{\mathop{{\bf {y}}}\nolimits}
\def\br{\mathop{{\bf {r}}}\nolimits}
\def\bs{\mathop{{\bf {s}}}\nolimits}
\def\bH{\mathop{{\bf {H}}}\nolimits}
\def\bk{\mathop{{\bf {k}}}\nolimits}
\def\be{\mathop{{\bf {e}}}\nolimits}
\def\bnul{\mathop{{\bf {0}}}\nolimits}
\def\bq{{\bf {q}}}

\newcommand{\oV}{\overline{V}}
\newcommand{\vkp}{\varkappa}
\newcommand{\os}{\overline{s}}
\newcommand{\opsi}{\overline{\psi}}
\newcommand{\ov}{\overline{v}}
\newcommand{\oW}{\overline{W}}
\newcommand{\oPhi}{\overline{\Phi}}

\def\mI{\mathop{{\mathfrak {I}}}\nolimits}
\def\mA{\mathop{{\mathfrak {A}}}\nolimits}

\def\st{\mathop{\rm st}\nolimits}
\def\tr{\mathop{\rm tr}\nolimits}
\def\sign{\mathop{\rm sign}\nolimits}
\def\d{\mathop{\rm d}\nolimits}
\def\const{\mathop{\rm const}\nolimits}
\def\O{\mathop{\rm O}\nolimits}
\def\Spin{\mathop{\rm Spin}\nolimits}
\def\exp{\mathop{\rm exp}\nolimits}

\def\mI{\mathop{{\mathfrak {I}}}\nolimits}
\def\mA{\mathop{{\mathfrak {A}}}\nolimits}

\def\st{\mathop{\rm st}\nolimits}
\def\tr{\mathop{\rm tr}\nolimits}
\def\sign{\mathop{\rm sign}\nolimits}
\def\d{\mathop{\rm d}\nolimits}
\def\const{\mathop{\rm const}\nolimits}
\def\O{\mathop{\rm O}\nolimits}
\def\Spin{\mathop{\rm Spin}\nolimits}
\def\exp{\mathop{\rm exp}\nolimits}

\title{Possible solution of the cosmological constant problem in the framework
of lattice quantum gravity}

\author {S.N. Vergeles\vspace*{4mm}\footnote{{e-mail:vergeles@itp.ac.ru}}}


\affiliation{Landau Institute for Theoretical Physics,
Russian Academy of Sciences,
Chernogolovka, Moskow region, 142432 Russia }

\begin{abstract} It is shown that in the theory of discrete quantum gravity the
cosmological constant problem can be solved due to the phenomena of
elliptic operators spectrum "loosening" and universe inflation.
\end{abstract}

\pacs{04.60.-m, 03.70.+k}

\maketitle

{\bf 1.} Let's outline shortly the cosmological constant problem
(see, for example, \cite{1}).

Consider Einstein equation with $\Lambda$-term ($\hbar=c=1$):
\begin{eqnarray}
R_{\mu\nu}-\frac12g_{\mu\nu}R=8\pi
G\,T_{\mu\nu}+\Lambda\,g_{\mu\nu}\,,
\label{introduction10}
\end{eqnarray}
\begin{eqnarray}
G\sim l_P^2\sim 2,5\cdot 10^{-66}\,{cm}^2\,.
\label{introduction20}
\end{eqnarray}
According to today experimental data
\begin{gather}
T_{\mu\nu}\sim 10^8{cm}^{-4}\longrightarrow
8\pi\,G\,T_{\mu\nu}\sim 5\cdot 10^{-57}{cm}^{-2}\,,
\label{introduction30}
\end{gather}
\begin{eqnarray}
\Lambda\sim 10^{-56}{cm}^{-2}\,.
 \label{introduction40}
\end{eqnarray}
Thus, if Einstein equation (\ref{introduction10}) is applied to
the today dynamics of universe, the quantities in
its right hand side are of the same order indicated in
(\ref{introduction30}) and (\ref{introduction40}).

On the other hand, any elementary estimation of the right hand side of
Eq. (\ref{introduction10}) in the framework of canonical quantum
field theory gives extremely large value in comparison with the experimental data
(\ref{introduction30}) and (\ref{introduction40}). Indeed, the vacuum expectation
value of the energy-momentum tensor in free quantum field theory has the order
\begin{eqnarray}
\langle T_{\mu\nu}\rangle_0\sim\int_{|{\bf k}|<k_{max}}\frac{\d^{(3)}k}{(2\pi)^3}
\left(\frac{k_{\mu}k_{\nu}}{k^0}\bigg|_{k^0=|\bk|}\right)\,.
\label{introduction60}
\end{eqnarray}
Here $k^{\mu}$ is the 4-momentum the corresponding mode.
If in the integral
(\ref{introduction60}) $k_{max}\sim l_P^{-1}$
(Planck scale), then it follows
from (\ref{introduction60}) and (\ref{introduction20})
\begin{eqnarray}
 8\pi G \langle T_{\mu\nu}\rangle_0\sim l_P^{-2}\sim
 10^{66}\,{cm}^{-2}\,.
 \label{introduction70}
\end{eqnarray}
It is clear that the interaction of fields doesn't changes
qualitatively the estimation (\ref{introduction70}) which in any case
is incompatible with the experimental estimations (\ref{introduction30}),
(\ref{introduction40}).

It is well known that the solution of the outlined problem is absent at present
\cite{1}, though some interesting ideas have been appeared lately (see, for example
\cite{2}, \cite{3}, \cite{4}). In this letter I do try to present the
qualitative estimations showing the compatibility of discrete quantum gravity
in quasi-classical state with the cosmological experimental data
(\ref{introduction30}) and (\ref{introduction40}). This means that in the considered
theory the vacuum expectation value of the energy-momentum tensor
becomes enough small due to (i) the phenomenon of "loosening"
of elliptic operators spectrum and (ii) inflation of universe. Moreover,
the quantum degrees of freedom exist right up to Planck scale.
The estimations are based on the results obtained in the papers \cite{4}, \cite{5},
\cite{6}.

{\bf 2.} It is necessary to write out some formulae from the work \cite{6}.
Here the notations are completely identical to that in \cite{6}.

Let $\mK$ be a 4-dimensional simplicial complex such that the
3-dimensional complex $\mS=\partial\mK$ has the topology of 3-sphere $S^3$.
To each vertex $a_i\in\mK$, the Dirac spinors
$\psi_i$ and $\overline{\psi}_i$ belonging to the
complex Grassman algebra are assigned.
To each oriented edge $a_ia_j\in\mK$, an element of the
group  $Spin(4)$
\begin{eqnarray}
\Omega_{ij}=\Omega^{-1}_{ji}=\exp\left(\frac{1}{2}\omega^{ab}_{ij}
\sigma^{ab}\right)\,, \ \ \ \sigma^{ab}=\frac{1}{4}[\gamma^a,\,\gamma^b]\,,
\label{discr20}
\end{eqnarray}
and also an element $\hat{e}_{ij}\equiv
e^a_{ij}\gamma^a$, such that
\begin{eqnarray}
\hat{e}_{ij}=-\Omega_{ij}\hat{e}_{ji}\Omega_{ij}^{-1}, \quad
-\infty<e^a_{ij}<\infty,
\label{discr30}
\end{eqnarray}
are assigned. The notations
$\overline{\psi}_{Ai}, \ \psi_{Ai}, \ \hat{e}_{Aij}, \
\Omega_{Aij}$ and so on indicate that the edge $X^A_{ij}=a_ia_j$ belongs to 4-simplex
with index $A$. Let $a_{Ai}, \ a_{Aj}, \ a_{Ak}, \ a_{Al}$, and $a_{Am}$ be all
five vertices of a 4-simplex with index $A$ and
$\varepsilon_{Aijklm}=\pm 1$ depending on whether the order of
vertices $a_{Ai}\,a_{Aj}\,a_{Ak}\,a_{Al}\,a_{Am}$ defines the
positive or negative orientation of this 4-simplex. The Euclidean action
of the theory has the form
\begin{eqnarray}&
I=\frac{1}{5\times
24}\sum_A\sum_{i,j,k,l,m}\varepsilon_{Aijklm}\tr\,\gamma^5 \times
\nonumber \\&
\times\left\{-\frac{1}{2\,l^2_P}\Omega_{Ami}\Omega_{Aij}\Omega_{Ajm}
\hat{e}_{Amk}\hat{e}_{Aml}+\right.
\\&
\left.+\frac{i}{48}\gamma^a\left(\overline{\psi}_{Ai}\gamma^a
\Omega_{Aij}\psi_{Aj}-\overline{\psi}_{Aj}\Omega_{Aji}\gamma^a\psi_{Ai}\right)
\hat{e}_{Amj}\hat{e}_{Amk}\hat{e}_{Aml}\right\}.
\nonumber
\label{discr40}
\end{eqnarray}

The oriented volume of a 4-simplex with vertexes
$a_{Ai}, \ a_{Aj}, \ a_{Ak}, \ a_{Al}$, and $a_{Am}$ is equal to
\begin{gather}
V_A=\frac{1}{4!}\frac{1}{5!}\sum_{i,j,k,l,m}\varepsilon_{A\,ijklm}\,
\varepsilon^{abcd}\,e^a_{A\,mi}e^b_{A\,mj}e^c_{A\,mk}e^d_{A\,ml}\,.
\label{discr80}
\end{gather}
The quantity
$l^2_{ij}=\sum_{a=1}^4(e^a_{ij})^2$
is interpreted as the square of the length of the edge $a_ia_j$.

The partition function $Z$ for a discrete
Euclidean gravity \footnote{I mean here that the partition
function is a functional of the values of the dynamic
variables at the boundary $\partial\mK$.} is defined as follows:
\begin{gather}
Z=\const\cdot\bigg (\prod_{\cE}\,\int\, \d\Omega_{\cE}\,\int\,\d
e_{\cE}\,\bigg)\times
\nonumber \\
\times\big(\prod_{\cV}\,\d\overline{\psi}_{\cV}\,
\d\psi_{\cV}\,\big)\,\exp\big(-I\,\big)\,. \label{discr280}
\end{gather}
Here, the vertices and edges are enumerated
by indices $\cV$ and $\cE$, and the
corresponding variables are denoted by $\psi_{\cV}$, \ $\Omega_{\cE}$,
respectively, $\d\Omega_{\cE}$
is the Haar measure on the group $\Spin(4)$, and
\begin{gather}
\d e_{\cE}\equiv\prod_a\,\d e^a_{\cE}\,, \quad \
\d\overline{\psi}_{\cV}\,\d\psi_{\cV}\equiv\prod_{\nu}\,
\d\overline{\psi}_{\cV\nu}\,\d\psi_{\cV\nu}\,. \label{discr290}
\end{gather}
The index $\nu$ enumerates individual components of the
spinors $\psi_{\cV}$ and $\overline{\psi}_{\cV}$.

Let's denote by ${\cal X}$ a 4-dimensional smooth manifold with topology of
the complex $\mK$. Consider a set of maps $\{g\}$ from geometrical realization
of the complex $\mK$ onto manifold ${\cal X}$ which
are not necessarily one-one maps. For a given local coordinates
$x^{\mu},\,\mu=1,\,2,\,3,\,4$ a map $g$ defines the coordinates
of images of vertexes $a_{A\,i}$: $x^{\mu}_{g(A\,i)}\equiv g^{\mu}(a_{A\,i})$.
Define the four differentials
\begin{gather}
\d x^{\mu}_{g(A\,ji)}\equiv x^{\mu}_{g(A\,i)}-x^{\mu}_{g(A\,j)}\,,
\quad i\neq j\,, \ \ \ i=1,\,\dots,4.
\label{discr140}
\end{gather}
Suppose the smooth fields $\omega_{\mu}^{ab}(x), e^a_{\mu}(x), \opsi(x),
\psi(x)$ are defined on the manifold ${\cal X}$. Then we can define
the discrete lattice variables according to the relations
\begin{gather}
\omega_{\mu}^{ab}(x_{g(A\,m)})\d x^{\mu}_{g(A\,mi)}=\omega_{A\,mi}^{ab},
\nonumber \\
e_{\mu}^a(x_{g(A\,m)})\d x^{\mu}_{g(A\,mi)}=e_{A\,mi}^a,
\quad \psi(x_{g(A\,i)})=\psi_{A\,i}.
\label{discr143}
\end{gather}
On the contrary, the discrete lattice variables in the right hand sides
of Eqs. (\ref{discr143}) define the values of the fields on the images
of vertexes of the complex. It is clear \cite{6} that for
the discrete lattice variables which change enough smoothly along the
complex we obtain the enough smooth fields. Moreover, in this case
the discrete action (\ref{discr40}) transforms to the well known
continuum action
\begin{eqnarray}&
I=\int\,\varepsilon_{abcd}\,\left\{-\frac{1}{l^2_P}R^{ab}\wedge e^c\wedge e^d+\right.
\nonumber \\&
\left.+\frac{i}{12}\,\big(\,\opsi\,\gamma^a\, {\cal
D}_{\mu}\psi-{\cal D}_{\mu}\opsi\,\gamma^a\,\psi\,\big)\,\d
x^{\mu}\wedge e^b\wedge e^c\wedge e^d\right\},
\nonumber \\&
e^a=e^a_{\mu}\d x^{\mu}, \qquad \omega^{ab}=\omega^{ab}_{\mu}\d x^{\mu},
\\&
\d\omega^{ab}+\omega^a_c\wedge\omega^{cb}=\frac{1}{2}\,R^{ab}.
\nonumber
\label{30}
\end{eqnarray}
I emphasize that we obtain the action (\ref{30}) only if the
lowest derivatives of the fields are taken into account. It is important
that in this case the information on the structure of the complex is lost.
This is incorrectly if the highest derivatives of the fields are
also taken into account. In the work \cite{6} the arguments have been given
that the quasi-classical phase at the same time is the macroscopic continuum phase with
long correlations and hence, also, the phase in which
the highest derivatives of the fields can be ignored. In this phase the partition
function is saturated by normal (smooth) modes but not by anomal modes responsible for
Wilson state doubling \cite{5}.

It is important that in the quasi-classical phase the universe wave function
does not depend on the discrete variables $e^a_{A\,ij}$ in a wide
diapason. This statement is true in the same context as the
highest derivatives of the fields can be ignored. Indeed, in the
quasi-classical phase the action (\ref{30}) as well as
the universe wave function depend on the fields
$\omega_{\mu}^{ab}(x), e^a_{\mu}(x), \opsi(x), \psi(x)$
which are present in the left hand side of Eqs. (\ref{discr143}).
Thus, fixing these fields and varying the maps $g$ or, equivalently,
the images of vertexes $x^{\mu}_{g(A\,i)}$
and, hence, the differentials $\d x^{\mu}_{g(A\,ji)}$, one can
vary the discrete variables $e_{A\,mi}^a$
in the right hand side of Eqs. (\ref{discr143}) in a wide region.

To clarify the situation, let's consider the case when
geometry of the space-time is flat or almost flat.
In the flat case one can take
\begin{gather}
\omega_{ij}^{ab}=0\,\,, \ \ \
\bigl(e_{ij}^a+e_{jk}^a+\ldots+e_{li}^a\bigr)=0\,\,.
\label{discr411}
\end{gather}
Here, the sum in the parentheses is taken over any closed path
formed by 1-simplices. Equations (\ref{discr411}) indicates that the
curvature and torsion are equal to zero. Thus, geometrical
realization of the complex $\mK$
is in the four-dimensional Euclidean space, $e^a_{ij}$ being the
components of the vector in a certain orthogonal basis in this
space, and if $R^a_i$ is the radius-vector of vertex $a_i$, then
$e_{ij}^a=R^a_j-R^a_i$. In this case one can take $e^a_{\mu}(x)=\delta^a_{\mu}$.
It is evident that Eqs. (\ref{discr411}) are the only restrictions for
the variables $e^a_{ij}$.

{\bf 3.} Now we are able to study the problem of spectrum loosening.
Here the spectrums of elliptic operators on the subcomplex $\partial\mK$
are studied. Note that although even in the simplest case
the subcomplex $\partial\mK$ has the topology of 3-sphere $S^3$,
here one can consider enough large subcomplexes of $\partial\mK$
as flat one and keep in mind Eqs. (\ref{discr411}).

Firstly, I write out the trivial formula for the volume in
momentum space occupied by all modes of scalar field defined on the vertexes
of periodic cubic lattice with the total number of vertexes $N$ and volume $V$
Here the spectrum of elliptic operators
on the subcomplex $\partial\mK$:
\begin{gather}
\Omega=(2\pi)^3\frac{N}{V}\,.
\label{discr420}
\end{gather}
The estimation (\ref{discr420}) remains true for all fields with the spin of the order of one;
moreover, this estimation remains qualitatively true for the case of irregular lattice
when the spacings between neighbor vertexes are commensurable.
I say that in this case the modes are densely packed in momentum space.

Now let's pass to the spectrum "loosening" problem in the discrete quantum
gravity.

The subsequent estimations in this Subsection are made for the intermediate
regime from confinement phase to the quasi-classical phase. In \cite{6}
the arguments are given that in this regime
the fields at nearest regions of space volume are
correlated weakly. It is natural to assume that the same conclusion
remains true at initial times in quasi-classical phase. Therefore
let us divide the macroscopic volume $V$ with the total number of
degrees of freedom (or the number of modes which in one's turn is of the order of the
number of vertexes of the complex) $N$ into ${\cal N}$
subvolumes $v_i$ in each of which $n_i$ degrees of freedom is
contained. Thus
\begin{gather}
\sum_{i=1}^{{\cal N}}n_i=N\,, \qquad  \sum_{i=1}^{{\cal N}}v_i=V\,,
\label{discr430}
\end{gather}
and
\begin{gather}
\omega_i=(2\pi)^3\frac{n_i}{v_i}
\label{discr440}
\end{gather}
is the minimal possible volume in momentum
space occupied by $n_i$ modes placed in the volume $v_i$.
Now instead of the quantity (\ref{discr420}) one must consider the
following quantity
\begin{gather}
\tilde{\Omega}=\frac{(2\pi)^3}{{\cal N}}\sum_{i=1}^{{\cal N}}\frac{n_i}{v_i}\,.
\label{discr450}
\end{gather}
Indeed, the minimum of quantity (\ref{discr450}) subjected to the
constraints (\ref{discr430}) is equal to (\ref{discr420}).

Since in the considered theory the volumes $v_i$ are variable
quantities, one must introduce the measure on the manifold of
volumes $\{v_i\}$. In \cite{6} the measure
\begin{gather}
\d\mu=\frac{({\cal N}-1)!}{V^{{\cal N}-1}}\delta\left(V-\sum_{i=1}^{{\cal N}}v_i\right)
\prod_{i=1}^{{\cal N}}\d v_i\,, \quad v_i>0\,,
\nonumber \\
\int\d\mu=1
\label{discr460}
\end{gather}
was suggested.

To justify the measure (\ref{discr460}), I note that the volumes and
the forms of elementary cells of the space are arbitrary and mutually independent,
and the wave function of universe depend weakly on these values in wide diapason
in the quasi-classical phase of the theory (see the
previous Subsection). The volumes of elementary cells
are determined according to Eq. (\ref{discr80})
only by 1-forms $e^a_{ij}$ which change independently
in integral (\ref{discr280}), and the action under
integral for partition function depends weakly (does not depend at all
on the discrete variables
satisfying Eqs. (\ref{discr411}) in the long-wave limit) on the variables $e^a_{ij}$ in
the quasi-classical phase. Moreover, the measure in integral (\ref{discr280})
is proportional to the product of
all differentials $\d e^a_{ij}$, which in one's turn is
proportional to the product $\prod_{i=1}^{{\cal N}}\d v_i$. So
we see that the simplest measure (\ref{discr460}) is valid for averaging.

Hence, instead of (\ref{discr450}) the more physically sensible
quantity is
\begin{gather}
\langle\tilde{\Omega}\,\rangle\equiv\int\tilde{\Omega}\d\mu=(2\pi)^3\frac{{\cal
N}-1}{V\,{\cal N}}\sum_{i=1}^{{\cal N}}n_i\int_{v_i\ll V}\frac{\d
v_i}{v_i}=
\nonumber \\
=(2\pi)^3\frac{N}{V}\int_{v_i\ll V}\frac{\d v_i}{v_i}\,.
 \label{discr470}
\end{gather}
The last equality follows from the first
constraint in (\ref{discr430}) and the relation ${\cal N}\gg 1$.

The comparison of Eqs. (\ref{discr420}) and (\ref{discr470}) shows
that the phenomenon of
essential expansion of the momentum space volume occupied by
quantum field modes arises as the consequence of the dynamics of the system.
This expansion factor is of the order of
\begin{gather}
\varkappa_1\sim\int_{v_i\ll V}\frac{\d
v_i}{v_i}=3\ln\frac{a_1}{a_0}=3\ln\xi_0\,.
\label{discr480}
\end{gather}
Here $a_0$ is some minimal dimension of the theory and $a_1\ll
V^{1/3}$. It seems that $a_0\gg l_P$, since only at $|e_{ij}^a|\gg l_P$
the quasi-classical phase can exist (see \cite{6}).

Now it is necessary to take into account that
instead of quantities $({n_i}/{v_i})$ in Eq. (\ref{discr450}) one must use the averaged (over
shortest wavelength fluctuations)
quantities $\langle({n_i}/{v_i})\rangle$.
Thus, using the obtained
estimation (\ref{discr480}) we elaborate a kind of
renormalization group describing loosening of mode packing.
Let $J$ be the number of steps of renormalization group and
\begin{gather}
\xi_j=\frac{a_{j+1}}{a_j}=\xi\gg 1\,, \quad j=1,\ldots\,,J\,,
\label{discr490}
\end{gather}
$a_{J+1}=a_{max}\lll a$, and $a$ is the universe radius. Thus
$\xi^J=\xi_1\xi_2\ldots\xi_J=a_{max}/a_0$. For rough estimation let us
take
\begin{gather}
J=\frac{1}{\lambda}\ln\frac{a_{max}}{a_0}\gg 1\,, \quad \ln\xi=\lambda\gg 1\,.
\label{discr500}
\end{gather}
Using Eqs. (\ref{discr480})--(\ref{discr500}) it is easy to see
that the expansion factor of momentum space volume occupied by
modes after $J$ steps is
\begin{gather}
\varkappa_J=\prod_{j=1}^J(3\ln\xi_j)=(3\ln\xi)^J=\left(\frac{a_{max}}{a_0}\right)^{(\ln
3\lambda)/\lambda}\,.
\label{discr510}
\end{gather}
The value of the right hand side of Eq. (\ref{discr510}) can bee very
large (many orders) in magnitude. This phenomenon is called here
as "spectrum loosening". It seems that the effect of spectrum
loosening and translational invariance are compatible on the
breathing lattice.

It follows from the presented analysis, that the continuum quantum
gravity arising from the discrete quantum gravity (if it exists)
possess very unusual properties. For example, let's try to estimate the
contribution to the cosmological constant due to the quantum field fluctuations
in the framework of presented here theory taking into account
the estimation (\ref{discr510}). We shall see that following this path
one can solve the problem of a large value of cosmological constant.
Indeed, in the elaborated here theory the estimation
(\ref{introduction70}) should be corrected by the factor $\varkappa_J^{-1}\lll 1$
 for the reason of noncompact
packing of the field modes in momentum space! Thus, instead of the estimation
(\ref{introduction70}) now we have the following one
\begin{gather}
\Lambda_{eff}\sim \left(\frac{a_0}{a_{max}}\right)^{(\ln
3\lambda)/\lambda}l_P^{-2}\lll l_P^{-2}\,.
\label{discr540}
\end{gather}

So the effective cosmological constant can be made enough small.

{\bf 4.}  In \cite{6} the arguments have been given that the
quasi-classical phase of the theory does actually exist.
Here in the Subsection 3 it is shown that
the properties of such theory are very unusual for the reason
of "spectrum loosening" phenomenon at the beginning of universe inflation.
Now the question arises: what looks like such unusual continuum quantum theory
of gravity? In this Subsection I try to describe phenomenologically a variant of
continuum quantum theory of gravity with the necessary properties.
Below under the dynamic system the continuum quantum theory of gravity
is meant. Here the results of the work \cite{4} are used. To quantize
the general covariant theories I follow the well known Dirac quantization
procedure. For clearness it is convenient to formulate the main assumptions
in the form of axioms.

{\bf {Axiom 1.}} {\it {All states of the theory having physical sense
are obtained from the ground state}} $\vert \,0\,\rangle$ {\it
{using the creation operators}} $A^{\dag}_n$:
 \begin{gather}
 \vert \,n_1;\ldots
 ;\,n_s\,\rangle=A^{\dag}_{n_1}\cdot\ldots\cdot
 A^{\dag}_{n_s}\,\vert \,0\,\rangle \ ,
\nonumber \\
   A_n\,\vert\,0\,\rangle=0 \,.
 \label{dq12}
\end{gather}
{\it {States (\ref{dq12}) form an orthogonal basis of the space of
physical states of the theory.}}

Here the operators $A^{\dag}_n$ and their conjugates $A_n$ are
the generators of bosonic and
fermionic Heisenberg algebra. For
the case of compact spaces which is interesting for us, the index $n$ belongs to a discrete
finite-dimensional lattice.

Since states (\ref{dq12}) are physical, they satisfy the relations
\begin{gather}
{\cal H}_T \, \vert \,n_1;\ldots;\,n_s\,\rangle=0\,\,,
\label{dq13}
\end{gather}
where ${\cal H}_T$ is the complete Hamiltonian of the theory. We
assume that ${\cal H}_T=\sum_{\Xi}v_{\Xi}\chi_{\Xi}$, where
$\{\chi_{\Xi}\}$ is the complete set of the first class
constraints and $\{v_{\Xi}\}$ is an arbitrary set of Lagrange
multipliers.

Equations (\ref{dq12}) and (\ref{dq13}) are compatible if and only if the
following relations are valid:
\begin{gather}
[A_n,\,{\cal
H}_T]=\sum_{\Xi,\,\Pi}r_{n\,\Xi\,\Pi}v_{\Xi}\,\chi_{\Pi}
\longleftrightarrow
\nonumber \\
\longleftrightarrow [A_n^{\dag},\,{\cal
H}_T]=-\sum_{\Xi,\,\Pi}\chi_{\Pi}\,v_{\Xi}^*r_{n\,\Xi\,\Pi}^{\dag}\,.
\label{dq15}
\end{gather}

Let an arbitrary field (or more general operator) $\Psi(x)$ be represented as a normal
ordered power series in operators $(A^{\dag}_n,\,A_n)$ (here
the index $n$ is fixed):
\begin{gather}
\Psi=\Psi'+\psi_n^{(+)}A_n+A^{\dag}_n\psi_n^{(-)}\,.
\label{dq16}
\end{gather}
Here $\psi_n^{(\pm)}$ are treated as the wave functions of the corresponding states.
By definition, here the operator $\Psi'$ does not depend on the
operators $(A^{\dag}_n,\,A_n)$:
\begin{gather}
[\Psi',\,A^{\dag}_n]=[\Psi',\,A_n]=0\,.
\label{dq17}
\end{gather}
It follows from Eqs. (\ref{dq15})--(\ref{dq17}) that
\begin{gather}
[\Psi,\,{\cal H}_T]=[\Psi',\,{\cal
H}'_T]+
\nonumber \\
+\sum_{\Xi}(q_{\Xi}\chi_{\Xi}+\chi_{\Xi}\tilde{q}_{\Xi})+(p_nA_n+A_n^{\dag}\tilde{p}_n)\,.
\label{dq18}
\end{gather}
Here the total Hamiltonian ${\cal H}_T$ is represented according
to (\ref{dq16}), so that ${\cal H}'_T$ does not depend on the operators
$(A^{\dag}_n,\,A_n)$.

Now let's impose an additional pair of second class constraints
\begin{gather}
A_n=0\,, \qquad A^{\dag}_n=0\,.
\label{dq23}
\end{gather}
By definition, under the constraints (\ref{dq23}) any operator $\Psi$ is
reduced to the operator $\Psi'$ in (\ref{dq16}). For any operators $\Psi,\,\Phi$
the Dirac bracket arising
under the constraints (\ref{dq23}) is defined according to the following
equality:
\begin{gather}
[\Psi,\,\Phi]^*\equiv[\Psi',\,\Phi']\,.
\label{dq24}
\end{gather}
The remarkable property of the considered theory is the fact that
\begin{gather}
[\Psi,\,{\cal H}_T]^*\approx [\Psi,\,{\cal H}_T]\,.
\label{dq25}
\end{gather}
Here the approximate equality means that after the imposition of
all first and second class constraints the operators in the both
sides of Eq. (\ref{dq25}) coincide, that is the weak equality
(\ref{dq25}) reduces to the strong one. Relation (\ref{dq25})
follows immediately from Eqs. (\ref{dq18}) and (\ref{dq24}). Eq.
(\ref{dq25}) means that the Heisenberg equation
\begin{gather}
i\dot{\Psi}=[\Psi,\,{\cal H}_T]^*
\label{dq28}
\end{gather}
for any field in reduced theory coincides weakly with
corresponding Heisenberg equation in nonreduced theory. Evidently,
this remarkable conclusion retains true under imposition of any
number of pairs of the second class constraints of the type
(\ref{dq23}). Thus, it is naturally to accept

{\bf {Axiom 2.}} {\it {In the case of compact space the index}} $n$
{\it {in axiom 1 runs a finite set of indexes:}} $n=1,\,\ldots\,,\,N$.
{\it {Moreover, it is assumed that the packing of modes in momentum
space is essentially noncompact.}}

{\bf{Axiom 3.}} {\it{The equations of motion and the constraints for
the physical fields have the
same form (up to the arrangement of the operators) as the
corresponding classical equations and constraints.}}

The axiom 2 states not only ultraviolet regularization of the theory
but also the "spectrum loosening" phenomenon. The axiom 3 is a consequence
of Eqs. (\ref{dq25}) and (\ref{dq28}).

In considered theory the totality of equations of motion and constraints
include Einstein equations and matter field equations of motion.
Further for brevity I shall call all these equations as equations of motion.
To obtain the solutions in such theory one must substitute to the equations of motion
the fields decomposed according to
\begin{gather}
\Psi(x)=\Psi'(x)+\sum_{n=1}^N\left\{\psi_n^{(+)}(x)A_n+A^{\dag}_n\psi_n^{(-)}(x)\right\}\,.
\label{dq30}
\end{gather}
By definition only the wave functions $\{\psi_n^{(\pm)}\}$ are decomposed in series of
operators $\{A_n^{\dag},\,A_n\}$ but not the field $\Psi'$. So the equations of motion become
the series in the powers of operators $\{A_n^{\dag},\,A_n\}$; evidently, the c-numerical
coefficients at different powers of operators $\{A_n^{\dag},\,A_n\}$ are
equal to zero separately. Thus the chain of c-numerical differential equation arises,
the zeroth approximation of which is the classical Einstein equation.
It is important that the equations of motions are general covariant.
This is the consequence of axiom 3 and the fact that the corresponding classical
equations are general covariant.

I call the quantization of gravity stated by axioms 1--3 as
dynamic quantization \cite{4}, \cite{7}.

{\bf 5.} Now, using the aforesaid, let's give some general estimations.

For simplicity let's consider the contribution to the cosmological constant
due to Dirac sea. The contribution to the energy-momentum tensor from the
massless Dirac field on the mass shell has the form
\begin{gather}
T_{\psi\,\mu\nu}=
\frac{i}{2}\Bigl(\overline{\psi}\,\gamma^a\,e_{a(\mu}
D_{\nu)}\psi-e_{a(\mu}\overline{D_{\nu )}\psi}\,\gamma^a\,\psi\Bigr),
\label{dqg33}
\end{gather}
and according to axioms 1 and 2 the Dirac field is decomposed as follows:
$$
\psi (x)=\sum_{n=1}^N \Bigl(a_n\,\psi_n^{(+)}(x)+b_n^{\dag}
\,\psi_n^{(-)}(x)\Bigr)+\ldots \,,
$$
$$
\{\,a_m\,,\,a_n^{\dag}\,\}=\{\,b_m\,,\,b_n^{\dag}\,\}=
\delta_{m,n}\,, \quad
 a_n|0\rangle=b_n|0\rangle=0\,.
$$
Here the positive/negative-frequency (in a sense, see \cite{7}) wave functions
$\{\psi_n^{(\pm)}(x)\}$
do not depend on $\{a_n,\,a_n^{\dag},\,b_n,\,b_n^{\dag}\}$.
It is easy to find the vacuum expectation
value of the quantity (\ref{dqg33}) in the second order \footnote{The order
of approximation means the total degree of operators
$\{A_n^{\dag},\,A_N\}$ taken into account.}:
\begin{gather}
\langle
T_{\psi\,\mu\nu}\rangle_0=\Re\left[i\sum_{n=1}^N\opsi_n^{(-)}\gamma^ae^{(0)}_{a(\mu}
D^{(0)}_{\nu)}\psi^{(-)}_n\right]\,.
\label{cc66}
\end{gather}

Now let us take into account that the dynamics of universe
is governed by the inflation scenario. In the zeroth
order the metric is expressed as
\begin{gather}
\d s^{(0)\,2}=\d t^2-a^2(t)\,\d\Omega^2\,,
\label{cc67}
\end{gather}
where $\d\Omega^2$ is the metric on unite sphere $S^3$ and $a(t)$
is the scale factor of universe. It follows from the Dirac equation
that the integrals
\begin{gather}
\int\d
V^{(0)}(t)\,\psi^{(\pm)\dag}_m(t)\psi^{(\pm)}_n(t)=\delta_{mn}
\label{cc68}
\end{gather}
are conserved. Using Eqs. (\ref{cc68}) we obtain the estimations
for the wave functions in (\ref{cc66}):
\begin{gather}
\left|\opsi_n^{(\pm)}(t)\,\psi_n^{(\pm)}(t)\right|\sim 1/a^3(t)\,.
\label{cc70}
\end{gather}
Therefore the estimation for the value (\ref{cc66}) is as follows:
\begin{gather}
\langle
T_{\psi\,\mu\nu}(t)\rangle_0\sim (N\,k_{max})/a^3(t).
\label{cc71}
\end{gather}
It is naturally to suppose that
$k_{max}\sim l^{-1}_P\sim 10^{33}{cm}^{-1}$, or $k_{max}\sim k_{SS}$
if supersymmetry is restored at $|{\bf k}|>k_{SS}\sim 10^4 GeV\sim
10^{18}cm^{-1}$.

To estimate the number $N$ in (\ref{cc71}) (the total number of
degrees of freedom)
I use the following formula:
\begin{gather}
N\sim\int^{k_{max}}\frac{\d^3k}{(\Delta k_{min})^3}
\left(\frac{\Delta k_{min}}{|\bk|}\right)^{\alpha}, \quad \alpha=1.
\label{cc78}
\end{gather}
The measure under the integral (\ref{cc78}) is Lorentz-invariant
and the value $\Delta k_{min}$ has the sense of the
nearest momenta difference modulo. For $\alpha=0$ and
$\Delta k_{min}=2\pi/a(t)$ the measure in (\ref{cc78})
coincides with the usual one for dense mode packing. I take
\begin{gather}
\Delta k_{min}\sim 10^{21}cm\sim 10^{-7}L\,,
\label{cc81}
\end{gather}
where $L=10^{28}cm$ (the dimension of observed part of Universe).

Thus from Eqs. (\ref{cc71})--(\ref{cc78}) an interesting inequality is obtained:
\begin{gather}
16\pi G\langle
T_{\mu\nu}\rangle_0\sim\frac{l_P^2\,k^3_{max}}{(\Delta k_{min})^2a^3(t_0)}\leq\Lambda\,.
\label{cc79}
\end{gather}
From here the
following estimation for the present dimension of universe is find:
\begin{gather}
a(t_0)\geq 10^{15}L\,.
\label{cc82}
\end{gather}
Analogously, in supersymmetric case
\begin{gather}
a(t_0)\geq L\,. \label{cc83}
\end{gather}

\begin{acknowledgments}

This work was
supported by RFBR No. 04-02-16970-a.

\end{acknowledgments}


\begin{thebibliography}{99}



\bibitem{1}
S. Weinberg, Rev. Mod. Phys. {\bf{61}}, 1 (1989).

\bibitem{2}
G. E. Volovik, E-print archives gr-qc/0101111; "The
Universe in a
 Helium Droplet", Clarendon Press. Oxford 2003.

 \bibitem{3}
S. Randjbar-Daemi, V. Rubakov, E-print archives
hep-th/0407176; V. P. Nair, S. Randjbar-Daemi, E-print archives
hep-th/0408063.

\bibitem{4}
S.N. Vergeles, hep-th/0411096.

\bibitem{5}
S.N. Vergeles, JETP Letters, {\bf{82}}, 617 (2005).

\bibitem{6}
S.N. Vergeles, Nucl. Phys. {\bf{B 735}}, 172 (2006); hep-th/0512137.

\bibitem{7}
S.N. Vergeles, JETP, {\bf{118}}, 996 (2000)





\end{thebibliography}
\end{document}